\documentstyle[prb,aps,preprint]{revtex}		
\begin{document}
\draft

\title{Multigrid Methods in Electronic Structure Calculations}

\author{E.\ L.\ Briggs, D.\ J.\ Sullivan, and J.\ Bernholc}
\address{Department of Physics,
North Carolina State University, Raleigh, North Carolina 27695-8202}

\maketitle
\begin{abstract}

We describe a set of techniques for performing large scale {\em ab initio\/}
calculations using multigrid accelerations and a real-space grid as a basis.
The multigrid methods provide effective convergence acceleration and
preconditioning on all length scales, thereby permitting efficient
calculations for ill-conditioned systems with long length scales or high
energy cut-offs. We discuss specific implementations of multigrid and
real-space algorithms for electronic structure calculations, including an
efficient multigrid-accelerated solver for Kohn-Sham equations, compact yet
accurate discretization schemes for the Kohn-Sham and Poisson equations,
optimized pseudo\-potentials for real-space calculations, efficacious
computation of ionic forces, and a complex-wavefunction implementation for
arbitrary sampling of the Brillioun zone.  A particular strength of a
real-space multigrid approach is its ready adaptability to massively parallel
computer architectures, and we present an implementation for
the Cray-T3D with essentially linear scaling of the execution time with the
number of processors. The method has been applied to a variety of periodic and
non-periodic systems, including disordered Si, a N impurity in diamond, AlN in
the wurtzite structure, and bulk Al.  The high accuracy of the atomic forces
allows for large step molecular dynamics; e.g., in a 1 ps simulation of Si at
1100 K with an ionic step of 80 a.u., the total energy was conserved within 27
$\mu$eV per atom.
\end{abstract}

\narrowtext

\section{Introduction}

Over the last several decades algorithmic advances coupled with the
development of high speed supercomputers, have made {\em ab initio\/} quantum
mechanical simulations possible for a wide range of physical systems. These
methods have been used to provide a theoretical framework for interpreting
experimental results and even to accurately predict the material properties
before experimental data were available. However, the calculations are
currently restricted to systems containing a few hundred
atoms.\cite{BigCalculations} This limitation is set by the available computer
power, and the scaling of the computational work with the number of atoms. One
of the most successful of the recent techniques is the Car-Parrinello
method\cite{CP} in which the electronic orbitals are expanded in plane-wave
basis functions and the resulting Hamiltonian is iteratively diagonalized.

The practical and efficient extension of {\em ab initio\/} quantum methods to
larger and more difficult systems may be accomplished by the refinement and
improvement of traditional methods or by the development of new techniques.
Although highly successful, traditional plane-wave methods encounter
considerable difficulties when they are applied to physical systems with large
length scales, or containing first-row or transition-metal atoms. These
difficulties may be partially eliminated by the use of preconditioned
conjugate-gradient techniques,\cite{TeterPayneAllan,AriasEtAl} optimized
pseudo\-potentials,\cite{Vanderbilt1990,RappeEtAl,LinEtAl,LiRabii}
augmented-wave methods,\cite{Bloechl} or plane waves in adaptive
coordinates.\cite{Gygi1992,Hamann1995} However, these methods are still
constrained by the use of a plane-wave basis set, and the necessity of
performing Fast Fourier Transforms (FFT) between the real and reciprocal
spaces. While FFT's may be implemented in a highly efficient manner on
traditional vector supercomputers, the current trend in supercomputer design
is massively parallel architectures. It is difficult to implement
efficient FFT algorithms on such machines, due to the required long range
communications.

Real-space methods are inherently local and therefore do not lead to a large
communication overhead. The scaling of several critical parts of large
calculations is improved from $O(N^2 \log N)$ in a plane-wave representation to
$O(N^2)$, where $N$ is the number of atoms. Furthermore, preconditioning and
convergence acceleration are most effectively carried out in real space
(q.v. Section III). A real-space formulation is also required for efficient
implementations of $O(N)$ electronic structure methods, in which the
computational work required scales linearly with the number of atoms. These
methods impose a localization constraint on the electronic
orbitals\cite{ON-Wave} or the electron charge density,\cite{ON-Density} which
eliminates the $O(N^3)$ orthogonalization step.

Orbital-based real-space approaches, {\em e.g.\/} atom-centered or floating
gaussians, are very well established. Recently, however, there has been
substantial interest in developing real-space orbital-independent methods,
which permit systematic studies of convergence in the spirit of plane-wave
methods. These methods include finite elements,\cite{WhiteTeterWilkins}
grids,\cite{BernholcYiSullivan,ChelikowskyEtAl,BSB,Beck,Weare,%
GygiGalli,ZumbachModineKaxiras} and wavelets.\cite{ChoEtAl,WeiChou1995}

The finite-element method was applied by White {\em et al.}
\cite{WhiteTeterWilkins} to one-electron systems. They used both
conjugate-gradient and multigrid acceleration\cite{Brandt} to find the
ground-state wavefunction. Two of the present authors
\cite{BernholcYiSullivan} used a basis with a high density of grid points in
the regions where the ions are located, and a lower density of points in the
vacuum regions, in conjunction with multigrid acceleration, to calculate the
electronic properties of atomic and diatomic systems. The core electrons were
explicitly included and nearly singular pseudo\-potentials were used.  The
non-uniform grid led to order of magnitude savings in the basis size and total
computational effort. The multigrid iterations, which provide automatic
preconditioning on all length scales, reduced the number of iterations needed
to converge the electronic wavefunctions by an order of magnitude in these
multi-length-scale systems. Weare {\em et al.}\cite{Weare} used a similar
method to solve for the ground state of H$_2^+$. Wavelet
bases\cite{ChoEtAl,WeiChou1995} were used to solve the LDA equations for atoms
and the O$_2$ molecule.  Chelikowsky {\em et al.}\cite{ChelikowskyEtAl} have used
high-order finite-difference methods and soft nonlocal pseudo\-potentials on
uniform grids to calculate the electronic structure, geometry, and short-time
dynamics of small Si clusters and of an isolated SO$_2$ molecule. Beck {\em et
al.}\cite{Beck} have used uniform grids and a smeared nuclear potential to
examine the energetics and structures of atoms and small molecules.  They employed
multigrid iteration techniques to improve the convergence rates of the
Kohn-Sham functional. Real-space grids in curvilinear coordinates have been
used by Gygi and Galli\cite{GygiGalli} to compute the properties of atoms and
CO$_2$, and Zumbach {\em et al.}\cite{ZumbachModineKaxiras} tested it on
O$_2$. Seitsonen {\em et al.}\cite{SeitsonenPuskaNieminen} used a uniform grid
approach with pseudo\-potentials and a conjugate gradients scheme to calculate
the electronic structure of P$_2$, and to study a positron trapped by a Cd
vacancy in CdTe.

In a previous communication\cite{BSB} the present authors have outlined a
multigrid-based approach suitable for large scale calculations, together with
a number of test applications. These included calculations for a vacancy in a
64-atom diamond supercell, an {\em isolated\/} $C_{60}$ molecule using
non-periodic boundary conditions, a highly elongated diamond supercell, and a
32-atom supercell of GaN that included the Ga $3d$ electrons in
valence. Uniformly spaced grids were used; in this case an effective
``cut-off'' may be defined, which is equal to that of the plane-wave
calculation that uses the same real-space grid for the FFT's. Whenever
feasible, the corresponding calculations were also carried out using
plane-wave techniques, and the two sets of results were in excellent agreement
with each other. This paper provides a comprehensive description of the
real-space multigrid method, and reports extensions to non-uniform grids,
non-cubic grids, and to molecular dynamics simulations with highly accurate
forces.

Several computational issues absent from plane-wave and orbital-based methods
arise when using a real-space grid approach. In the plane-wave basis the
action of the kinetic energy operator on the basis functions can be computed
exactly, and the wavefunctions, potentials, and the electron charge density
can be trivially expanded in the basis. Every basis-set integral, except those involving
the LDA exchange-correlation functional, can be computed exactly.  The
computational errors in the calculations are mainly due to the truncation of
the basis. In a real-space grid implementation, the Kohn-Sham equations must
be discretized explicitly, which presents important trade-offs between
accuracy and computational efficiency. Furthermore, the quantum-mechanical
operators are known only at a discrete set of grid points, which can introduce
a spurious systematic dependence of the Kohn-Sham eigenvalues, the total
energy, and the ionic forces on the relative position of the atoms and the
grid. We have developed a set of techniques that overcome these difficulties
and have been used to compute accurate static and dynamical properties of
large physical systems, while taking advantage of the rapid convergence rates
afforded by multigrid methods.

This paper is organized as follows: In Section II a method for the accurate
and efficient real-space discretization of the Kohn-Sham equations for cubic,
orthorhombic, and hexagonal symmetries is described. Section III focuses on
the multigrid algorithms, which greatly accelerate convergence of the
electronic wavefunctions and of the Hartree potential. Tests of the
convergence acceleration are described in Section IV. The calculation of ionic
forces that are sufficiently accurate for large step molecular dynamics
requires special methods, which are described in Section V. Section VI
discusses performance issues for massively parallel supercomputers, and
describes a highly scalable and efficient implementation on the
Cray-T3D, which has been tested on up to 512 processors.
The summary in Section VII is followed by several technical
appendices.

\section{Grid-Based Discretizations of the Kohn-Sham Equations}

Electronic structure calculations that use a real-space mesh to represent the
wavefunctions, charge density, and ionic pseudo\-potentials must address a new set of
technical difficulties when compared with plane-wave methods.  In a plane wave
representation the form of the kinetic energy operator is obvious. In contrast, the
representation of the kinetic energy operator on a real-space grid is
approximated by some type of finite differencing, the accuracy of which must
be carefully tested. Below, we describe real-space discretizations for uniform
cubic, orthorhombic, and hexagonal grids, as well as nonuniform scaled cubic
grids that increase the resolution locally.  For uniform cubic grids we
also describe and test the extension to periodic systems with arbitrary
sampling of the Brillioun zone.

\subsection{Uniform Cubic and Orthorhombic Grids}

Previously,\cite{BSB} we described a real-space approach that uses uniform
cubic grids with $\Gamma$ point k-space sampling. In direct comparisons with
plane-wave calculations, we found nearly perfect agreement between the two
methods for several test systems. We now provide further details of our
method.

In real-space the wavefunctions, the electron charge density, and the
potentials are directly represented on a uniform three-dimensional real-space
grid of $N_{grid}$ points with linear spacing $h_{grid}$. The physical
coordinates of each point are
\begin{eqnarray}
{\bf r}(i,j,k) = &&  (i\, h_{grid}, j\, h_{grid}, k\, h_{grid}) \nonumber\\
i=1,\cdots,N_x; \ \ j= &&1,\cdots,N_y; \ \ k=1,\cdots,N_z.
\label{GRID1}
\end{eqnarray}
The ions are described by norm-conserving
pseudo\-potentials\cite{Hamann,BHS,HSC} in the Kleinman-Bylander nonlocal
form.\cite{KleinmanBylander} These potentials are interpolated onto
the grid from their radial representation. Exchange and correlation effects
are treated using the local density approximation (LDA) of density functional
theory, in which the total electronic energy of a system of electrons and ions
may be expressed as
\begin{eqnarray}
E_{LDA} =
&&
   \sum_{n=1}^{N_{states}} f_n \epsilon_n + \int d{\bf r} \rho({\bf r})
   \{ \epsilon_{XC}[\rho({\bf r})] -
           \mu_{XC}[\rho({\bf r})] \} \nonumber\\
&&
   -{1 \over 2} \int \, d{\bf r} \, \rho({\bf r}) \,
                        V_{Hartree}({\bf r}) + E_{ion-ion}.
\label{LDA-Energy}
\end{eqnarray}
The minimization of this functional requires the solution of the Kohn-Sham
equations
\begin{equation}
H_{KS}[\psi_n] =
    -{1 \over 2}\nabla^2 \psi_n + V_{eff} \psi_n = \epsilon_n \psi_n,
\label{KSequations}
\end{equation}
subject to the orthonormality constraint on the eigenfunctions
$<\psi_i|\psi_j> = \delta_{ij}$. The accurate discretization of these
equations on the grid structure described by Eq.\ (\ref{GRID1}) requires
appropriate numerical representations of the integral and differential
operators.  All integrations are performed using the three-dimensional
trapezoidal rule:
\begin{equation}
\int d{\bf r} f({\bf r}) \doteq h_{grid}^3 \sum_{ijk} f({\bf r}(i,j,k)).
\label{Integration}
\end{equation}
We have
found that for high accuracy it is essential that the integrand $f({\bf r})$
be band-limited in the sense that its Fourier transform should have minimal
magnitude in the frequency range $G > G_{max} \equiv \pi/h_{grid}$.  This is
explicit in a plane-wave calculation since the basis is cut-off at a specific
plane-wave energy.

The discrete real-space grid also provides a kinetic-energy cut-off of
approximately $G_{max}^2/2$.  Unlike the plane-wave basis, high-frequency
components above this cut-off can nonetheless manifest themselves on the grid.
This high-frequency behavior, which can introduce unphysical variation in the
total energy or the electron charge density, is perhaps best seen when the
ions, and hence their pseudo\-potentials, shift relative to the grid
points.\cite{BSB} If the pseudo\-potentials contain significant high-frequency
components near or above $G_{max}$ then, as the ions shift, the high frequency
components are aliased to lower frequency components in an unpredictable
manner.

This effect can be decreased by {\em explicitly\/} eliminating the
high-frequency components in the pseudo\-potentials by Fourier filtering.  In the
context of plane-wave calculations, King-Smith {\em et
al.}\cite{King-SmithPayneLin} recognized that the real-space integration of
the nonlocal pseudo\-potentials could differ significantly from the exact
result computed in momentum space, unless the potentials were modified so that
Fourier components near $G_{max}$ were removed. Fourier filtering of
the pseudo\-potentials is thus required in real-space calculations for accurate results.
It is, of course, possible to use unfiltered potentials on real-space grids
provided the grid spacing is sufficiently small, but our experience shows that
the total energy and the electron charge density are often sufficiently well
converged for significantly larger grid spacings --- provided that explicit
pseudo\-potential filtering is used. We use a somewhat different Fourier filtering
method than that proposed by King-Smith {\em et al.}, but it produces
essentially the same effect (see Appendix B).

The differential operator in the Kohn-Sham equations is approximated using a
{\em generalized\/} eigenvalue form:
\begin{eqnarray}
{\bf H}_{mehr}[\psi_n] = && \nonumber\\
{1 \over 2} {\bf A}_{mehr}[\psi_n] + {\bf B}_{mehr}[V_{eff} \psi_n] = &&
           \epsilon_n {\bf B}_{mehr}[\psi_n],
\label{Compact-Implicit}
\end{eqnarray}
where ${\bf A}_{mehr}$ and ${\bf B}_{mehr}$ are the components of the {\em
Mehrstellen\/} discretization,\cite{Collatz} which is based on Hermite's
generalization of Taylor's theorem. It uses a weighted sum of the wavefunction
and potential values to improve the accuracy of the discretization of the {\em
entire\/} differential equation, not just the kinetic energy operator.  In
contrast to the central finite-differencing method, this discretization uses
more {\em local\/} information (next-nearest neighbors, for example). The
definition of the fourth-order {\em Mehrstellen\/} operator used in the
present work is specified by the weights listed in Table
\ref{Mehrstellen-Ortho}, which pertain to both cubic and orthorhombic grids
(see below). A more detailed analysis of the {\em Mehrstellen\/} operator and
Eq.\ (\ref{Compact-Implicit}) is given in Appendix A.

This representation of the Kohn-Sham Hamiltonian is short-ranged in real space
in the sense that the operator can be applied to any orbital in $O(N_{grid})$
operations.  Specifically, the application of the ${\bf A}_{mehr}$ operator at
a point involves a sum over 19 orbital values while the application of the
${\bf B}_{mehr}$ operator requires a sum over 7 points. The local potential
multiplies the orbital pointwise, and the short-ranged nonlocal projectors
require one integration over a fixed volume around each ion and a pointwise
multiplication.  This sparseness permits the use of iterative diagonalization
techniques, and the short-ranged representation of the Hamiltonian leads to an
efficient implementation on massively parallel computers.

The discussion up to this point has been restricted to uniform cubic grids,
but the extension to a general orthorhombic grid is straightforward. There are
now three separate grid spacings $h_x$, $h_y$, and $h_z$ with the coordinates
of each grid point given by:
\begin{eqnarray}
{\bf r}(i,j,k) = &&  (i\, h_x, j\, h_y, k\, h_z) \nonumber\\
i=1,\cdots,N_x; \ \ j= &&1,\cdots,N_y; \ \ k=1,\cdots,N_z.
\label{GRID1a}
\end{eqnarray}
The orthorhombic {\em Mehrstellen\/} operator described in Table
\ref{Mehrstellen-Ortho} is used to discretize the Kohn-Sham equations and
numerical integration is performed according to
\begin{equation}
\int d{\bf r} f({\bf r}) \doteq h_x \, h_y \, h_z \,
                                \sum_{ijk} f[{\bf r}(i,j,k)].
\label{Integration1}
\end{equation}

\subsection{Extension to Arbitrary Bloch Wave-Vectors}

In the preceding Section the wavefunctions were assumed to be real, with
the Brillioun-zone sampling restricted to the $\Gamma$ point. When these
restrictions are lifted the Kohn-Sham equations become
\begin{eqnarray}
-{1 \over 2}\nabla^2 \psi_{nk} + && i{\bf k} \cdot {\bf \nabla} \psi_{nk} +
     {1 \over 2} |{\bf k}|^2 \psi_{nk} + \nonumber\\
 && V_{eff}({\bf k}) \psi_{nk} = \epsilon_n \psi_{nk},
\label{CKSequations}
\end{eqnarray}
\noindent
where the $\psi_{nk}$ are now the periodic parts of the Bloch functions.  They
are complex-valued, which presents no additional difficulties in
discretization.  The nonlocal projectors in $V_{eff}({\bf k})$ have been
multiplied by the phase factor $e^{i {\bf k} \cdot {\bf r}}$. The gradient
term ${\bf \nabla} \psi_{nk}$ is computed using a central finite difference
expression, which in one dimension has the form:
\begin{equation}
{d \over dx} f(x_i) = \sum_{n=-3}^3 \alpha_n f(x_{i+n}) + O(h^7),
\end{equation}
where $\alpha_1 = 3/(4 h_{grid})$, $\alpha_2 = -3/(20 h_{grid})$, $\alpha_3 =
1/(60 h_{grid})$, and $\alpha_{-n} = -\alpha_n$. For a cubic grid structure
the three dimensional generalization of this is the sum of the individual
expressions for each coordinate axis. Denoting this finite-difference operator
by $\tilde {\bf \nabla}$, the discretization of the Kohn-Sham equation becomes
\begin{eqnarray}
{\bf H}_{bloch}[\psi_{nk}] && = {1 \over 2} {\bf A}_{mehr}[\psi_{nk}] + \nonumber\\
    && {\bf B}_{mehr}[i{\bf k} \cdot \tilde {\bf \nabla} \psi_{nk} +
        {1 \over 2} {\bf k}^2\psi_{nk} + V_{eff}({\bf k}) \psi_{nk}] \nonumber\\
   && = \epsilon_n {\bf B}_{mehr}[\psi_{nk}],
\label{ComplexCompact-Implicit}
\end{eqnarray}
where ${\bf A}_{mehr}$ and ${\bf B}_{mehr}$ are again the components of the
{\em Mehrstellen\/} operator.

The accuracy of the discretization was tested by calculating the lattice
constant and bulk modulus for an 8-atom Si supercell.  A 26-Ry equivalent
cut-off was used with k-space sampling restricted to the Baldereschi
point.\cite{Baldereschi} The calculated lattice constant was 5.38 \AA\ with a
bulk modulus of 0.922 Mbar. These are in good agreement with the corresponding
plane-wave calculation with the cut-off of 26 Ry: 5.39 \AA\ and 0.960 Mbar,
respectively.\cite{BuongiornoNardelli}

Bulk aluminum was selected as an additional test case. A 4-atom cell was used
with a 23 Ry equivalent cut-off and k-space sampling of 35 special points in
the irreducible part of the Brillouin zone.  Since Al has partially occupied
orbitals, the bands near the Fermi level were occupied using a Fermi-Dirac
broadening function of width 0.1 eV. The convergence of the total energy with
respect to energy cut-off and the number of k-points was tested by increasing
the cut-off from 23 to 49 Ry, which produced a change of only 10 meV/atom, and
by increasing the number of k-points to 56, which changed the total energy by
3 meV/atom.  The calculated lattice constant and bulk modulus of 4.02 \AA\ and
0.734 Mbar are in excellent agreement with the experimental values of 4.02
\AA\ and 0.722 Mbar, and with previous theoretical results of Lam and
Cohen\cite{LamCohen} of 4.01 \AA\ and 0.715 Mbar.

\subsection{Uniform Hexagonal Grids}

Unlike plane-wave methods, where different symmetry groups can be handled
easily, an {\em efficient\/} real-space implementation for periodic systems
with non-orthogonal lattice translation vectors requires considerable
modifications to the orthorhombic-symmetry implementation.

The hexagonal grid describing the unit cell or supercell is generated by
\begin{eqnarray}
&& {\bf r}(i,j,k) = h_{xy} i\, {\bf n_1} +
                    h_{xy} j\, {\bf n_2} + h_z k\, {\bf n_3} \nonumber\\
&&i=1,\cdots,N_x; \ \ j= 1,\cdots,N_y; \ \ k=1,\cdots,N_z,
\label{HexGrid}
\end{eqnarray}
where the ${\bf n_i}$ are the hexagonal Bravais-lattice
vectors.\cite{AshcroftMermin} The $c/a$ ratio can be chosen arbitrarily by
varying the two independent grid spacings, $h_{xy}$ and $h_z$, and the number
of grid points, $N_x = N_y$ and $N_z$.  However, one should use $h_{xy} \sim
h_z$ in order to maintain similar resolution both in the {\em xy} plane and
along the {\em z}-axis.  Because the indexing of this hexagonal grid is
isomorphic to the cubic one, the computer representation of potentials and
wavefunctions does not change. The most important difference is in the
discretization of the Kohn-Sham equations. We have implemented a mixed
sixth-order kinetic energy operator. This discretization is described in
Appendix C, as well as the modifications required in the multigrid restriction
and interpolation procedures (q.v. Section III).

The above implementation has been tested on a 32-atom supercell of AlN in the
wurtzite phase. The accuracy of the results was confirmed by comparison with
plane-wave Car-Parrinello calculations on the same supercell.  Generalized
norm-conserving\cite{Hamann,BHS,HSC} Al and N pseudo\-potentials were used for
both calculations with k-space sampling restricted to the $\Gamma$ point. The
cohesive energy from the real-space calculation was 11.5 eV per AlN unit,
which compares well with the value of 11.6 eV obtained in the Car-Parrinello
calculations. The eigenvalue degeneracies were identical in both calculations
and the maximum difference in any eigenvalue was 0.04 eV.

The extension of the real-space grid representation to other Bravais lattices
proceeds in a similar manner, the only requirement being the existence of an
accurate finite difference discretization.

\subsection{Scaled Grids}

In real-space calculations it is possible to add resolution locally. This is
especially valuable for systems with a wide range of length scales such as
surface or cluster calculations. A high density of grid points can be used
near the ions, with a low density in the vacuum regions. Other possible
applications are simulations of impurities in bulk materials, where the
impurity ions may require higher resolutions to be accurately represented. By
using locally enhanced regions the required resolution may be added only where
needed, thereby greatly reducing the total number of points required.

Local enhancement of the grid resolution may be achieved by adding small
high-resolution grids\cite{BernholcYiSullivan,Weare} onto a uniform global
grid, or by using a coordinate transform to warp the grid structure. Our focus
here will be on the second approach, which was first proposed by
Gygi\cite{Gygi1992} for plane-wave basis sets and recently extended to real
space by several workers.\cite{GygiGalli,ZumbachModineKaxiras} In the
real-space approach, a continuous coordinate transform is applied to a uniform
grid. In general the transformation is non-separable, but we prefer a
separable coordinate transformation in order to avoid mixed derivatives in the
kinetic-energy operator. As a test of the utility and accuracy of this
scaled-grid approach, we examined an interstitial oxygen impurity in
Si\cite{NeedelsEtAl} for two grid layouts: a dense uniform grid with a 76-Ry
cut-off and a scaled grid with a cut-off that varied from 18 Ry to 76 Ry. The
scaling-transformation that maps the fictitious computational grid $x$ to the
physical warped grid $x_{scaled}$ is
\begin{equation}
x_{scaled}(x) =
   ( x-x_0) - { L_x \beta\over 2\pi}\sin({ 2\pi (x - x_0) \over L_x}),
\end{equation}
where $x_0$ is the $x$-coordinate of the oxygen atom, $L_x$ is the length of
the supercell in the $x$-direction, and $\beta$ is an adjustable parameter
between zero and one that controls the degree of scaling. The $y$ and $z$
coordinates are scaled analogously.  The scaled grid required four times fewer
points than the uniform grid to achieve the same convergence of the total
energy.

Since the coordinate transforms are continuous functions, the integration
weights and the coefficients of the discretized kinetic-energy operator may be
generated from the uniform grid values using the metric tensor of the
transform in the manner outlined by Gygi.\cite{Gygi1992,GygiGalli} With these
modifications, the calculations proceed as for the uniform grids, but a
sixth-order central finite-difference operator is used to discretize the first
and second derivatives because we have not constructed a {\em Mehrstellen\/}
operator for the scaled grid. In each case (uniform and scaled grids), a
64-atom supercell was used and the silicon atoms were fully relaxed. The
oxygen atom was held fixed in order to avoid Pulay corrections\cite{Pulay} to
the ionic forces, which would have been required if it and the scaled grid
were free to move. The uniform and scaled-grid calculations are in very good
agreement: the maximum difference in Kohn-Sham eigenvalues was 40 meV and the
maximum difference in ionic coordinates was 0.03 \AA.

\section{Multigrid Algorithms}

To efficiently solve Eq.\ (\ref{Compact-Implicit}), we have used
multigrid-iteration techniques that accelerate convergence by employing a
sequence of grids of varying resolutions. The solution is obtained on a grid
fine enough to accurately represent the pseudo\-potentials and the electronic
wavefunctions. If the solution error is expanded in a Fourier series, it may
be shown that iterations on any given grid level will quickly reduce the
components of the error with wavelengths comparable to the grid spacing but
are ineffective in reducing the components with wavelengths large relative to
the grid spacing.\cite{Brandt,MGTut} The solution is to treat the lower
frequency components on a sequence of auxiliary grids with progressively
larger grid spacings, where the remaining errors appear as high frequency
components. This procedure provides excellent preconditioning for all length
scales present in a system and leads to very rapid convergence rates. The
operation count to converge one wavefunction with a fixed potential is
$O(N_{grid})$, compared to $O(N_{grid}\,\log N_{grid})$ for FFT-based
approaches.\cite{PayneEtAl}

There is no one multigrid algorithm but rather a collection of algorithms that
share certain common features. In order to describe the implementation used in
this work, we start with a description of a multigrid solver for Poisson's
equation. This will then be used as a building block for the more
sophisticated algorithms actually employed.

A standard numerical problem that illustrates the multigrid algorithm is the
Poisson equation, $-\nabla^2 V_{Hartree} = 4\pi\rho$, defined on a rectangular
cell of dimensions $(L_x, L_y, L_z)$ with periodic boundary conditions. A
standard method of solving for $V_{hartree}$ is to discretize the equation on
a uniform three dimensional grid with spacing $h_{grid}$ and $N_{grid}$ total
points. The differential operator $-\nabla^2$ is represented by some form of
finite differencing. This produces a set of linear algebraic equations
\begin{eqnarray}
{\bf A}{\bf x} = {\bf b},
\label{MG1}
\end{eqnarray}
where ${\bf x}$ and ${\bf b}$ are the discretized forms of $V_{Hartree}$ and
$4\pi\rho$, respectively, and ${\bf A}$ is the finite-difference
representation of $-\nabla^2$. If the system is small, direct matrix methods
are an acceptable means of solving the equations; however, for large systems
the work required scales as $(N_{grid})^3$, which is prohibitive.  An
alternative approach is to use an {\em iterative\/} relaxation scheme such as
the Jacobi method.\cite{Recipes1} In this technique, the solution is
iteratively improved.  First, define $\tilde {\bf x}$ as an approximate
solution of Eq.\ (\ref{MG1}), and the residual ${\bf r}$, a measure of the
solution error, as
\begin{eqnarray}
{\bf r} = {\bf b} - {\bf A} \tilde {\bf x}.
\label{RES1}
\end{eqnarray}
An improved $\tilde {\bf x}$ is generated using
\begin{eqnarray}
\tilde {\bf x}^{new} = \tilde {\bf x} + \Delta tK {\bf r},
\label{RES2}
\end{eqnarray}
where ${\Delta}t$ is a pseudo time step and K is the inverse of the diagonal
component of ${\bf A}$. This approach will always converge to a solution for
some value of ${\Delta}t$ if ${\bf A}$ is diagonally
dominant.\cite{Numerical1} However, the number of iterations required to
reduce the magnitude of the residual to a specified accuracy is proportional
to $N_{grid}^{2/3}$, so that the algorithmic cost to converge,
$O(N_{grid}^{5/3})$, is too great.\cite{Recipes1} While more sophisticated
relaxation methods such as Gauss-Seidel, successive overrelaxation (SOR), or
the alternating direction implicit method (ADI)\cite{Recipes1} have improved
convergence rates, they still scale as $(N_{grid})^a$ with $a > 1$, and are
too slow for the grid sizes required in electronic structure calculations.

The slow convergence of the Jacobi method can be qualitatively understood by
noting that because ${\bf A}$ is a short-ranged operator, the updated
approximate solution, Eq.\ (\ref{RES2}), is a linear combination of nearby
values. If the error in the current estimate of $\tilde {\bf x}$ is decomposed
into Fourier components, it can be shown that one Jacobi iteration
considerably reduces the high frequency components of the error, but many
Jacobi iterations are needed to affect the longest wavelength components of
the error.  The overall convergence rate is then limited by that of the lowest
frequency components. As the problem becomes larger, the lowest frequency
representable on the grid becomes smaller and the convergence rate decreases.

The essence of the multigrid approach is the observation that the individual
frequency components of the error are best reduced on a grid where the
resolution is of the same order of magnitude as the wavelength of the error
component. This approach will maintain a high convergence rate for all
frequency components of the error even when the problem size (and the grid)
becomes very large.

We first describe a multigrid algorithm to solve Poisson's equation that uses
two grids; a fine grid of spacing $h$ and a coarser auxiliary one of spacing
$H$.  In this work, we use a coarse-to-fine grid ratio $H/h$ of 2, but other
ratios are possible.  The solution is generated as follows: the high-frequency
components of the solution error with wavelength $\approx~h$ are reduced by
one or two Jacobi iterations.  The residual ${\bf r}_h$, which should be
devoid of high-frequency variation, is computed and transferred to the coarse
grid by restriction (see below).  Next, Poisson's equation on the coarse grid
with the residual as a source term is solved by using the same iteration
procedure as in Eq.\ (\ref{RES2}), but with an initial estimate of zero. The
Jacobi iteration on this level removes error components with wavelength
$\approx~H$.  Finally, the coarse grid solution is interpolated to the fine
grid and added to the fine grid solution.  This process is referred to as a
coarse-grid-correction scheme (CGC). A few applications of the CGC cycle are
generally sufficient to solve Poisson's equation to machine precision even for
extremely large systems.

An obvious question is how the solution is obtained on the coarse grid. If the
total number of grid points in the coarse grid is small, a direct matrix
method will be sufficient. If this number is so large that a direct method is
impractical, then a second, coarser grid level is introduced and the two-grid
algorithm is repeated in a recursive manner. When multiple grid levels are used,
the pattern of cycling through the grids also needs to be considered. We use a
simple progression from the finest to the coarsest grid level and then back to
the finest level, which is referred to as a V-cycle. More complicated cycling
schemes exist, but we have found that the V-cycle works as well in
electronic structure calculations as the more sophisticated approaches.
Another consequence of the multigrid approach is the reduction in the size of
the grid (and consequently in the work required) on each level. For a uniform
grid in three dimensions a doubling of $h_{grid}$ with each level leads to a
factor of eight reduction in $N_{grid}$, so that the addition of extra coarse
grid levels is computationally inexpensive.

The simplest choice for the restriction operator is to copy every other point
in the fine grid directly to the coarse grid. This so-called straight
injection, while easy to implement, does not always yield good convergence
rates. A better choice is a weighted restriction, in which each coarse-grid
value is the average of the 27 fine-grid values surrounding it. In our work,
the weight assigned to each fine-grid point is proportional to the volume it
occupies at a given coarse-grid point.  A good choice for interpolating
from the coarse grid to the fine grid is the adjoint of the restriction
operator, which in this case becomes simple tri-linear
interpolation.\cite{Recipes2}

The final accuracy of the solution is determined by the finite-difference
representation of $-\nabla^2$ on the finest grid level. It is neither
necessary nor desirable to use the same representation of $-\nabla^2$ on all
grid levels; i.e., ${\bf A}_h$ may differ from ${\bf A}_H$ in the form of the
discretization as well as in the grid spacing. The technique of changing
discretization on different levels is referred to as deferred defect
correction (DDC) or double discretization.\cite{Brandt,Hackbusch} It is
especially valuable for problems where the operator used on the finest grid
level is numerically unstable on the coarser grids, or when it is inconvenient
to apply. The accuracy of the fine-grid solution does not depend on the choice
of this coarse-grid operator, which effects only the convergence rate.  In
solving Poisson's equation, we use the {\em Mehrstellen\/} operator on the
finest grid level for high accuracy. However, it is unsuitable for convergence
acceleration on the coarser grids because of stability problems. Thus, on the
coarser grids a 7-point central finite-difference operator is used, which
provides excellent stability and rapid high-frequency attenuation.

The extension of multigrid concepts to the solution of the Kohn-Sham equations
introduces several complications.  First, the equations are non-linear since
the eigenvalues and the orbitals must be computed simultaneously.  Second,
when the Kleinman-Bylander form of the nonlocal pseudo\-potentials is used,
the equations become a set of integro-differential eigenvalue equations.
Finally, the Hamiltonian depends upon the density, and must be solved
self-consistently.

Brandt {\em et al.}\cite{BrandtMcCormickRuge} have given a multigrid
algorithm for the standard eigenvalue problem.  Below we describe
an alternative means of linearizing the eigenvalue equations.  The multigrid
technique recommended in the literature for non-linear integro-differential
equations is the full approximation storage (FAS) method.\cite{Brandt} In FAS
the {\em entire\/} problem is discretized and solved on {\em all\/} grid
levels. In contrast, the CGC method outlined above generates the full solution
only on the finest level. While the theoretical performance of FAS on this
problem is superior, its implementation is significantly more complex.
In addition, it
is difficult to obtain an accurate representation of the nonlocal
pseudo\-potentials on the coarser grid levels.  Furthermore, the Kohn-Sham
equations need not be converged to maximum accuracy at every iteration, because
the electron charge density (and therefore the Hamiltonian) change after each
multigrid step.

For these reasons, a modification of the double discretization approach
described above was used. The discretized operator on the finest grid is the
{\em Mehrstellen\/} approximation of the Kohn-Sham Hamiltonian, while a
7-point central finite-difference representation of $-\nabla^2$ alone is used
on the coarse grids. Effectively, the coarse-grid equation for the
wavefunction residual becomes Poisson's equation, where the source term is the
residual generated by the {\em Mehrstellen\/} operator on the fine grid, Eq.\
(\ref{MehrRes}).
 
Our multigrid procedure begins with the selection of some initial
wavefunctions and electron charge density. We postpone discussion of
initialization techniques until later and assume that an adequate start has
been generated. The following steps are then performed for each individual
wavefunction: First, an estimate of the eigenvalue is calculated from the
Rayleigh quotient of the generalized eigenvalue equation, Eq.\
(\ref{Compact-Implicit}):
\begin{equation}
\epsilon_n =
   { \langle\psi_n| {\bf H}_{mehr} [\psi_n]\rangle
      \over
     \langle\psi_n| {\bf B}_{mehr} [\psi_n]\rangle }.
\label{RD1a}
\end{equation}
In the case of complex orbitals, the eigenvalue is calculated using the
Rayleigh quotient of the real (or imaginary) part of the generalized
eigenvalue equation, Eq.\ (\ref{ComplexCompact-Implicit}):
\begin{equation}
\epsilon_n =
   { \langle Re[\psi_{nk}]| Re[{\bf H}_{bloch} [\psi_{nk}]]\rangle
      \over
     \langle Re[\psi_{nk}]| Re[{\bf B}_{mehr} [\psi_{nk}]]\rangle }.
\label{RD1b}
\end{equation}
Next, several Jacobi iterations are applied to the orbital on the finest grid
using Eqs.\ (\ref{RES1}) and (\ref{RES2}), where the residual is computed as
\begin{equation}
{\bf r}_h = \epsilon_n {\bf B}_{mehr}[\psi_n] - {\bf H}_{mehr}[\psi_n].
\label{MehrRes}
\end{equation}
The fictitious time step $\Delta t$ used in the Jacobi iteration is typically
chosen between $0.8$ and $1.4$ a.u.  In the case of complex orbitals, the real
and imaginary components of the orbital are updated separately, using the
appropriate generalization of Eq.\ (\ref{MehrRes}).

Next, the residual is restricted to the first coarse grid.  A DDC coarse-grid
cycle begins using the 7-point central finite-difference representation of
$-\nabla^2$ instead of $H_{mehr}$.  Several auxiliary coarse grids can be
used.  When the coarse-grid correction is interpolated onto the finest grid,
only a fraction $\beta_{CGC}$ of it is added to the orbital, for reasons of
stability. A value of $\beta_{CGC} = 0.5$ has been found to work for almost
all systems. (Larger values may produce much higher convergence rates on some
systems while being unstable for others, so some experimentation is
necessary.)

Before transferring the residual to the coarse grid, it is essential that
enough Jacobi iterations be performed to eliminate the high frequency
components from the residual. Since the residual is used as the right hand
side of a CGC correction cycle, any high frequency components will eventually
be transferred to a coarser grid where they cannot be represented correctly,
greatly reducing the effectiveness of the multigrid cycle. In some cases they
may even make the process numerically unstable.  In the above approach, the
difficulties of discretizing the nonlocal pseudo\-potentials on the coarse grid
levels are avoided because the potential term is computed on the finest grid
and frozen thereafter.

The steps outlined above in the DDC apply only to a single wavefunction. The
full solution process also requires the application of the orthonormality
constraints and an update of the electron charge density. The full solution
process (one SCF step) consists of the following cycle.

First, the DDC is applied to all of the wavefunctions.  Next, the
orthonormality constraints are applied using the Gram-Schmidt procedure:
\begin{eqnarray}
\tilde \psi_i = && \psi_i - \sum_{j < i} \psi_j^{new}\,
                    \langle \psi_j^{new} | \psi_i \rangle, \nonumber \\
\psi_i^{new}  = && \tilde \psi_i /
            \sqrt{\langle\tilde\psi_i|\tilde\psi_i\rangle}, \nonumber \\
            i = && 1, \cdots, N_{states}.
\label{SCF1}
\end{eqnarray}
The new electron charge density is generated by linear mixing:
\begin{equation}
\rho_{new} = (1 - \alpha ) \rho_{old} +
             \alpha \sum_{i=1}^{N_{states}} f_i \psi_i^2,
\label{SCF2}
\end{equation}
where $f_i$ is the occupation of the $i^{\hbox{th}}$ state and $\alpha$ is a
mixing parameter, generally set to a value between 0.5 and 0.9.  Next, the
Hartree potential is recomputed for the new charge density using a {\em
Mehrstellen\/} DDC cycle, and a new exchange-correlation potential is
generated.

Finally, a subspace diagonalization may be performed at this point. This need
only be done occasionally (every 10-20 SCF steps is generally adequate) in
order to unmix eigenstates that may be close in energy.  Because the {\em
Mehrstellen\/} Hamiltonian leads to a non-hermitian generalized eigenvalue
equation (see Appendix A), subspace diagonalization requires a brief
discussion: We look for a unitary transformation of the current wavefunctions
that better represents the eigenvectors of the Hamiltonian, and are led to
the following eigenvalue equation for the subspace:
\begin{equation}
  \sum_n H^{sub}_{m,n} d_{n,l} = \epsilon_l \sum_n B^{sub}_{m, n} d_{n, l},
\label{subspace}
\end{equation}
where
\begin{eqnarray}
 H^{sub}_{m,n} = && \langle\psi_m| {\bf H}_{mehr} [\psi_n]\rangle \\
 B^{sub}_{m,n} = && \langle\psi_m| {\bf B}_{mehr} [\psi_n]\rangle,
\end{eqnarray}
and $d_{n, l}$ is the matrix of coefficients of the unitary transformation for
the $l^{\hbox{th}}$ state.  Because $B^{sub}_{m, n}$ is invertible (see
Appendix A), the subspace equations are equivalent to
\begin{equation}
   \sum_n C^{sub}_{m,n} d_{n,l} = \epsilon_l d_{m, l},
\label{subspace2}
\end{equation}
where $C^{sub} = (B^{sub})^{-1}H^{sub}$.  The matrix $C^{sub}$ is not
hermitian except when the subspace is a subset of the space of eigenvectors.
Thus, we do {\em not\/} diagonalize $C^{sub}$ because its eigenvectors are not
necessarily orthogonal, which would spoil the orthogonality of the electronic
orbitals.  Instead, we discard the anti-hermitian part of $C^{sub}$, which is smaller
than the hermitian part of $C^{sub}$ by $O(h_{grid}^2)$, and
diagonalize the hermitian part.  This approximation works well in practice,
and is exact at convergence.

The hermitian approximation does not affect the final accuracy of the solution
because the multigrid-assisted Jacobi iterations ultimately converge the
orbitals.  Nonetheless, accurate subspace rotations are essential for good
convergence: compare the convergence rates in Figs.\
\ref{SiliconConvergence}-\ref{DiamondConvergence}. As a test, we compared
subspace diagonalizations with the {\em Mehrstellen\/} and the hermitian
sixth-order discretizations, and found that the convergence is significantly
improved with the former.

The cycle described above is repeated until the electronic system converges to
the desired tolerance, which may be monitored by computing the RMS value of
the residual vector for each wavefunction (see Eq.\ (\ref{MehrRes})). When
this reaches a value of $10^{-9}$ a.u.\ for all wavefunctions in the occupied
subspace, the convergence is sufficient for the computation of forces that are
accurate enough for large step molecular dynamics with excellent energy
conservation.

As was mentioned previously, the convergence rates depend on the choice of the
initial wavefunctions and electron charge density. A poor choice can lead to
slow convergence rates or in some cases the system will not converge at
all. Apart from random initial wavefunctions or an approximate solution that
is generated using an LCAO basis set, one can also use a double-grid
scheme. In the latter method the initial solutions are generated on a grid
with a spacing twice as large as that used for the final grid. The
computational work on this coarse grid is eight times smaller than what is
needed on the fine grid. The approximate coarse-grids wavefunctions are
then interpolated to the fine grid and used as the initial guess. This process can
reduce the number of SCF cycles needed on the finest grid level by a factor of
two to three, thereby achieving significant savings in the computational
effort.

\section{Tests of Multigrid Convergence Acceleration}

The theoretical convergence rates of multigrid methods may, in principle, be
calculated exactly for certain types of problems. For well-behaved partial
differential equations such as Poisson's equation discretized on $N_{grid}$
points, $O(N_{grid})$ total operations are required to obtain a solution
accurate to the the grid-truncation error. This compares well with FFT based
methods which require $O[N_{grid}\log(N_{grid})]$ operations. For the
Kohn-Sham equations, an exact theoretical bound on multigrid convergence rates
is difficult to obtain due to self-consistency effects, and to the best of our
knowledge this analysis does not yet exist. We have therefore elected to study
convergence properties in an empirical fashion by performing tests on physical
systems typical of the problems normally examined with density functional
theory.

In previous work\cite{BSB} the present authors examined convergence rates for
8-atom supercells of perfect diamond as a function of the effective
kinetic-energy cut-off determined by the grid resolution, for a 32-atom
supercell of GaN that included the Ga $3d$ electrons in valence, and for a highly
elongated 96-atom diamond supercell. It was found that multigrid convergence
rates were largely independent of energy cut-off and cell geometries. While
promising, these results were obtained for perfect crystal configurations of
semiconductor compounds, which are generally fairly easy to converge. In this
article we present the results of a more systematic study that includes
disordered systems.

The first system selected was a 64-atom supercell of bulk silicon. The ions
were represented by a generalized norm-conserving
pseudo\-potential\cite{Hamann,BHS,HSC} and the grid spacing used corresponded
to an energy cut-off of 12 Ry. The ionic positions were generated by a
molecular dynamics simulation at a temperature of 1000 K. Because the work
required to converge to the ground state depends on the quality of the initial
wavefunctions and charge density, we used random initial wavefunctions and a
constant initial electron charge density to minimize any possible bias from
the choice of a starting configuration. A small number (10\% of the total) of
conduction band states was included in the calculations. Fig.\
\ref{SiliconConvergence} shows the convergence rate defined as the
$\log_{10}(E-E_0)$, plotted as a function of iteration number, where each
iteration represents a single SCF step. Results are shown for calculations
performed with and without multigrid acceleration, where the latter used a
steepest-descents algorithm. In addition, the two calculations were repeated
with, and without subspace diagonalizations of the orbitals. For the
calculations that included subspace diagonalizations, the procedure was
applied every 8 SCF steps, which led to small discontinuities in the smooth
evolution of the total energy. The results show that maximum convergence rates
are obtained when multigrid iterations are combined with subspace
diagonalization.  The slowest convergence occurs for steepest descents with no
subspace diagonalization. For the two runs where subspace diagonalizations
were performed, the multigrid run converged at roughly 2.5 times the rate of
the steepest descents approach.

While these results are encouraging, bulk silicon is a relatively
straight-forward test and is well handled by standard plane-wave methods. As
an example of a more difficult system we have considered a 64-atom diamond
supercell with a substitutional nitrogen impurity. Standard
pseudo\-potentials\cite{Hamann,BHS,HSC} were used for both C and N. The strong
N $p$ potential required an energy cut-off of 63 Ry. The presence of a
localized nitrogen donor level together with the 63 Ry cut-off makes the
system more difficult to converge. Random initial wavefunctions were used, and
Fig.\ \ref{DiamondConvergence} shows the observed convergence rates.  The
convergence rates for the two runs that use subspace diagonalization are a
factor of 4 better for multigrid than for steepest descents. This relative improvement
is considerably greater than that observed for the silicon cell and is the
consequence of the automatic preconditioning provided by multigrid techniques
for all of the length and energy scales present in the problem. The multigrid
convergence rates are largely independent of the grid spacing, which roughly
corresponds to the kinetic energy cut-off in plane-wave approaches. This is
not true of the steepest descents algorithm, where the maximum stable time
step that may be used decreases as the energy cut-off increases.

When comparing the convergence rates of the multigrid and steepest descents
approaches, the computational workload involved in each technique must also be
considered. A particular advantage of multigrid methods, when compared to
other convergence acceleration schemes, is their low computational cost. This
is due to the factor of 8 reduction in the number of grid points on each
 successive
multigrid level. The computational time per SCF step in the silicon and
diamond runs described above increased by less than 10\% when multigrid was
used instead of steepest descents. For bigger systems, where the costs of
orthogonalizing the orbitals and applying the nonlocal pseudo\-potentials
begin to dominate the total computational time, the extra work needed for the
multigrid accelerations becomes negligible. In terms of computational
time, the 64-atom Si supercell described above required 1.6 seconds per SCF
step on 64 processors of a Cray-T3D.

\section{Ionic Forces and Molecular Dynamics}

Efficient structure optimizations and the calculation of dynamical quantities
such as phonon frequencies and thermodynamic properties require accurate ionic
forces. In plane-wave methods the ionic forces are computed by applying the
Hellmann-Feynman theorem.\cite{Hellmann,Feynman} Since the derivative of the
pseudo\-potentials may be expressed exactly within the plane-wave basis, the
accuracy of the ionic forces is limited only by machine precision and the
degree of convergence to the Born-Oppenheimer surface.

For the grid-based approach the accuracy of Hellmann-Feynman forces is limited
by the numerical error in computing the integrals of the derivatives
of the pseudo\-potentials.  This error decreases with grid spacing.  The
differentiation of the radial potentials and projectors must be performed with
care to include the effects of the Fourier filtering.  Alternatively, a
derivative-free implementation of the Hellmann-Feynman forces can be used,
which we term virtual displacements.  In this scheme, the ionic pseudo\-potentials are
numerically differentiated directly on the real-space grid.  The ions are
moved through a set of small displacements, while the electron charge density
and the wavefunctions are held fixed. The potential energy is calculated for each
displacement and finite-differenced to form the derivative.  The forces
computed by the two methods agree well.  Most of the forces we have calculated to
date have been computed using virtual displacements.

A stringent test of the accuracy of the ionic forces is a constant-energy
molecular dynamics simulation. Over the course of the simulation any
systematic errors in the forces will manifest themselves as poor energy
conservation. A distinction has to be made between small random errors that
appear as bounded oscillations in the total energy and errors that increase in
magnitude with simulation time. The small random errors are expected in the
real-space approach because the energy of an ion varies by a small amount as
its position changes relative to the grid points.\cite{BSB} This is of no
particular concern as long as the magnitude of the variation is small and
oscillatory in nature. Of greater concern are errors that are unbounded.
These could arise from errors in the forces, errors in
integrating the equations of motion of the ions, and lack of self-consistency
due to inadequate convergence of the electronic wavefunctions. The first
source of error was minimized by Fourier filtering of the ionic
pseudo\-potentials. The second is generally not a problem unless the ionic
time step is too large. For small time steps even a simple integrator such as
the Verlet algorithm is sufficient, and larger time steps may be handled by
using higher order integrators, such as the Beeman-Verlet method.\cite{Beeman}
The last source of error is the most significant because Hellmann-Feynman
forces are only accurate to first order in the error of the wavefunctions. A
high degree of self-consistency is thus necessary to obtain good energy
conservation.

A 64-atom silicon supercell was selected to test energy conservation on a
typical system. The ions were given random initial displacements from the
perfect crystal configuration, and several velocity rescaling steps were
performed in order to attain an average ionic temperature of 1100 K. A
constant-energy molecular-dynamics simulation over 1 ps was then carried out,
using 80 a.u.\ time steps and third-order Beeman-Verlet\cite{Beeman}
integration of the ionic equations of motion. The potential, kinetic, and
total energy of the system vs. simulation time are plotted in Fig.\
\ref{simd}.  We observed good energy conservation: the maximum variation in
the total energy was 1.75 meV, which corresponds to 27 $\mu$eV per atom.

\section{Massively Parallel Implementation}

The performance of a given algorithm when solving complicated problems
depends not only on the theoretical efficiency, which may be quite high, but
also on how adaptable the algorithm is to modern computer architectures. One
example are certain classical molecular dynamics algorithms, which perform
only slightly better on vector supercomputers than on low cost engineering
workstations, even though the supercomputer's theoretical peak performance may
be an order of magnitude larger. A particular strength of the Car-Parrinello
method has been its efficient implementation on vector supercomputers, such as
the Cray-YMP. However, vector performance, while improving steadily, is
unlikely to increase by several orders of magnitude per decade as has occurred
in the past. At the same time, the development of powerful, low cost
microprocessors and memory, has led to massively parallel architectures
consisting of a large number of microprocessors, linked by a high speed
communication network. Although efficient implementations of plane-wave-based
methods on massively parallel architectures exist, the FFT-based algorithms do
not scale well with the number of processors because the FFT is a global
operation.

Below, we will describe a massively parallel implementation of the multigrid
method. Although some of the code-optimization issues are
architecture-specific, most are generic and thus applicable to any massively
parallel computation. The target machine is the Cray-T3D, which uses up to
2048 DEC-Alpha microprocessors, each with a peak performance of 150 MFlops.
Each processor has 8 KB direct-mapped data and instruction caches and 8 MW of
local memory. The processors are linked together in a three-dimensional torus
arrangement for data communication. Three issues have to be addressed in order
to write an efficient code for this type of machine: minimizing communication
costs between processors, balancing the work-load on each processor, and code
optimization on the individual processors.

\subsection{Data Decomposition and Load Balancing}

The majority of the data storage in the multigrid method consists of the
wavefunction values on the real-space grid. We will consider the case where
the points are distributed on a uniform three-dimensional rectangular grid. If
$N_{wf}$ is the total number of wavefunctions, then $N_{grid}N_{wf}$ total
storage is required. The simplest possible decomposition of data is to store
complete wavefunctions on each processing element (PE), where each PE stores
$N_{wf}/N_{PE}$ orbitals. While conceptually simple, this approach will
perform poorly for large systems with many wavefunctions, because
orthogonalizing wavefunctions residing on different PE's requires sending
large amounts of data between processors.  An alternative approach, and the
one adopted by us, is to use real-space data decomposition. In this method,
each PE is mapped to a specific region of space. The electron charge density,
Hartree potential, and each wavefunction are distributed by regions over the
processors.  With this approach interprocessor communication is restricted to
two areas: the computation of integrals on the real space grid (See Eq.\
(\ref{Integration})), and the application of the finite-differencing
operators.

For integration, the ideal optimization strategy is to organize the
calculation so that as many integrals as possible are computed at once.  This
can be understood by considering the time required for interprocessor
communication, which consists of a latency period and a transfer phase. The
latency period is significant and is the same whether 1 or 1000 words of data
is transferred.
Our integration procedure is as follows: First, calculate the
intra-processor contributions to the integral (i.e., integrate over the
subdomains); Second, store as many of these local integrals as possible;
Finally, transfer them between processors in blocks and complete the
integration by summing the local integrals.

It was straightforward to implement the above procedure in most cases, but the
orthogonalization step required significant modifications. In a standard
implementation of the Gram-Schmidt orthogonalization algorithm, wavefunction
overlaps and updates are computed sequentially, and the integrals cannot be
computed in parallel. To reduce the number of
data transfers, the following implementation of Gram-Schmidt orthogonalization
was adopted.  First, the overlap matrix $S_{ij} = \langle\psi_i|\psi_j\rangle$
is computed as above: the local parts of the overlap integrals are computed and
stored on each processor; the integration is then completed by transferring
them in blocks to the other processors.  Second, the Cholesky
factorization\cite{GolubVanLoan1} of the overlap matrix is computed: $S_{ij} =
(C^\dagger C)_{ij}$.  The Cholesky factor, $C$, is relevant because its
components are the overlaps between the new orthogonal wavefunctions and the
original ones.\cite{Sullivan1995} Finally, the diagonal components of the
Cholesky factor are used to normalize the wavefunctions, and the off-diagonal
ones are used to complete the orthogonalization:
\begin{eqnarray}
\psi_i^{new} = && {1 \over C_{i,i}}
           (\psi_i - \sum_{j < i} \psi_j^{new}\, C_{j,i})  \nonumber \\
           i = && 1, \cdots, N_{states}.
\label{NewGram}
\end{eqnarray}
For simplicity, the Cholesky factorization of $S_{ij}$ is currently performed
on each processor.  The computational time to factorize scales as
$(N_{wf}^3)$, but has not yet become a bottleneck.  However, for very large
systems (greater than 800 orbitals), a parallelized Cholesky
factorization will save significant computer time and memory.

The second area where interprocessor communication is required is the finite
differencing of the wavefunctions and Hartree potential, since finite
differencing is nonlocal. However, in the {\em Mehrstellen\/} discretization,
the nonlocality is restricted to points within one grid unit in each Cartesian
direction.  Interprocessor communication is thus always limited to nearest
neighbor PE's regardless of the size of the system.  This low communication
cost is a particular advantage of a {\em Mehrstellen\/} type approach as
opposed to a central finite-difference approach, which requires a higher
degree of nonlocality to achieve the same level of accuracy.

Load balancing and the efficient use of all PE's is a major concern for any
parallel algorithm. With the method described above, the load balancing is
essentially perfect for all parts of the calculation except for the
application of the nonlocal pseudo\-potentials. These are applied to the
wavefunctions in localized volumes around each ion. If the distribution of
ions in space is nonuniform, then some of the PE's will be idle for a fraction
of each SCF step. However, actual calculations on many systems have shown that
the application of the nonlocal potentials typically requires less than 10\%
of the total computational time on any PE, so that processor utilization will
always exceed 90\%.

The efficiency of the massively parallel implementation described here is
illustrated in Fig.\ \ref{speedup}, which shows the speedup in execution time
per step for a given problem as the number of PE's is increased. The graph
indicates a superlinear relationship, which is an artifact due to single
processor cache effects.  There are two competing factors here. The first is
the increased communication cost as the number of processors increases, which
tends to reduce the speedup. The second is the reduction in the amount of data
stored on each processor and a consequent increase in the number of cache
hits. As was discussed earlier, the communication costs are relatively small
with the data model being used; since the cache is relatively small, cache hit
effects outweigh these. An apparent superlinear speedup is observed.

\section{Summary}

We have described the development of a multigrid-based method that uses a
real-space grid as a basis. The multigrid techniques provide preconditioning
and convergence acceleration at all length scales, and therefore lead to
particularly efficient algorithms. A specific implementation of multigrid
methodology in the context of density functional theory was described and
illustrated with several applications. The salient points of our
implementation include: (i) the development of new compact discretization
schemes in real space for systems with cubic, orthorhombic, and hexagonal
symmetry, and (ii) the development of new multilevel algorithms for the
iterative solution of Kohn-Sham and Poisson equations.  The accuracy of the
discretizations was tested by direct comparison with plane-wave calculations
when possible and were found to be in excellent agreement in all cases.  These
algorithms are very suitable for use on massively parallel computers and in
$O(N)$ methods.  We described an implementation on the Cray-T3D massively
parallel computer that led to a linear speedup in the calculations with the
number of processors.

The above methodology was tested on a large number of systems. A prior
Communication described tests on C$_{60}$ molecule, and diamond and GaN
supercells. The present article examined convergence properties in detail for
a supercell of disordered Si and the N impurity in diamond. The multigrid
techniques increased the convergence rates by factors of 2 and 4,
respectively, when compared to the steepest descents algorithm. An extension
to non-uniform grids that uses a separable coordinate transform to change grid
resolution locally, e.g., at the surface or near an impurity, was developed
and tested on the O interstitial in Si. This extension results in only minor
changes in methodology and coding, while the reduction in basis set size and
thus in computational effort can be significant. A complex version of
multigrid code, capable of an arbitrary sampling of the Brillouin zone, was also
was developed and tested on bulk Al.

Large time-step molecular dynamics simulations require very accurate forces, which
can potentially lead to difficulties in real space methods as the atoms move
relative to the grid points. We have described a set of techniques based on
Fourier filtering of pseudo\-potentials that eliminate these difficulties for
grid spacings of sizes similar to those used in plane-wave calculations. A 1
ps test simulation of bulk Si at 1100 K conserved the total energy to within
27 $\mu$eV per atom, and illustrated the high  quality of these forces. Further
applications of this methodology are in progress, including a simulation of
surface melting of Si,\cite{WensellBAPS} structural properties of large
biomolecules that contain over 400 atoms,\cite{BrabecBAPS} and electronic and
structural properties of In$_x$Ga$_{1-x}$N quantum wells.\cite{BuongiornoNardelliBAPS} The
multigrid methodology is also very suitable for $O(N)$ implementations, and
tests results for a 216-atom cell of bulk Si were described
recently.\cite{SullivanBAPS}

\section{Acknowledgments}

The authors wish to thank C.\ J.\ Brabec for designing some of the parallel
multigrid routines; M.\ Buongiorno Nardelli for his help with the
complex-wavefunction and Brillioun zone sampling routines; and M.\ G.\ Wensell for
supplying his molecular dynamics routines.  This work was supported by ONR
grant number N00014-91-J-1516, NSF grant number DMR-9408437, and ONR grant
number N00014-96-1-0161. The calculations were performed at
the Pittsburgh Supercomputer Center and the North Carolina Supercomputing
Center.

\appendix

\section{Analysis of the {\em Mehrstellen\/} Operator}

The {\em Mehrstellen\/} discretization differs from central
finite-differencing in two important respects: first, higher accuracy in the
discretization is achieved by using more local information, but this accuracy
is fully realized only at convergence; and second, the discretized Kohn-Sham
eigenvalue equation Eq.\ (\ref{Compact-Implicit}) is non-hermitian because the
operator ${\bf B}_{mehr}$ does not commute with the potential operator.  In
this appendix, we examine the accuracy of the {\em Mehrstellen\/}
discretization, and prove that the non-hermitian nature of ${\bf H}_{mehr}$
does not change the nature of the wavefunctions: they remain orthogonal.  For
simplicity, we analyze only the {\em Mehrstellen\/} discretization of the
orthorhombic lattice.

The fourth-order {\em Mehrstellen\/} discretization (see Table
\ref{Mehrstellen-Ortho}) samples the Hamilton and the wavefunction at 19 points
\begin{eqnarray}
{\bf A}_{mehr}[f({\bf x})] =
  && a f({\bf x}) + \sum_{n=1}^3 b_n f({\bf x} \pm h_n \hat {\bf x}_n )
 \nonumber\\
  && + \sum_{n<m} c_{n,m} f({\bf x} \pm h_n \hat {\bf x}_n
                                    \pm h_m \hat {\bf x}_m )\\
{\bf B}_{mehr}[f({\bf x})] = &&
    a' f({\bf x}) + \sum_{n=1}^3 b'_n f({\bf x} \pm h_n \hat {\bf x}_n ).
\label{AppA-cubic}
\end{eqnarray}
The accuracy of the {\em Mehrstellen\/} discretization is one order higher
than the corresponding central finite-differencing one, but this accuracy is
achieved only at convergence.\cite{Collatz} The small $h$ expansions of the
${\bf A}_{mehr}$ and ${\bf B}_{mehr}$ demonstrate this principle:
\begin{eqnarray}
{\bf A}_{mehr} = && -\nabla^2 - {1\over 12}\nabla^2
                \sum_{n=1}^3 h_n^2 {\bf \nabla}_n^2 + O(h^4)\\
{\bf B}_{mehr} = && I + {1\over 12}
                \sum_{n=1}^3 h_n^2 {\bf \nabla}_n^2 + O(h^4)
\label{AppA-1}
\end{eqnarray}
Note that by construction, ${\bf A}_{mehr} = {\bf B}_{mehr} (-\nabla^2)$ to
$O(h^4)$.  Thus, the {\em Mehrstellen\/} discretization of the Kohn-Sham
equations is equivalent to
\begin{eqnarray}
&& {\bf H}_{mehr}[\psi_n] - {\bf B}_{mehr}[ \epsilon_n \psi_n] = \nonumber\\
&& {\bf B}_{mehr}[{\bf H}_{KS}\psi_n -\epsilon_n\psi_n] + O(h^4).
\label{AppA-2}
\end{eqnarray}
The $O(h^2)$ terms, implicit in the right-hand side, vanish at convergence, when ${\bf H}_{KS}\psi_n =
\epsilon_n\psi_n$.  A similar analysis applies to the discretization of the
Poisson equation:
\begin{eqnarray}
&& {\bf A}_{mehr}[V_H] - {\bf B}_{mehr}[4\pi\rho] = \nonumber\\
&& {\bf B}_{mehr}[-\nabla^2 V_H - 4\pi \rho] + O(h^4).
\label{AppA-3}
\end{eqnarray}

Unlike a plane-wave or central finite-differencing representation of the
Kohn-Sham equations, the {\em Mehrstellen\/} discretization Eq.\
(\ref{Compact-Implicit}) leads to a {\em non}-hermitian, {\em generalized\/}
eigenvalue equation. Nonetheless, we prove that the right eigenvectors of the
discretized operator, {\em i.e.\/} the electronic orbitals, are orthogonal
because they are also eigenvectors of a {\em hermitian\/} Hamiltonian.
The generalized eigenvalue equation can be recast into a more familiar form
by multiplication by ${\bf B}_{mehr}^{-1}$ (The invertibility of
${\bf B}_{mehr}$ is discussed below):
\begin{eqnarray}
{1 \over 2} ({\bf B}_{mehr}^{-1} {\bf A}_{mehr}) \psi_n + V_{eff} \psi_n && =
                   \nonumber\\
 {1 \over 2} {\bf C} \psi_n + V_{eff} \psi_n && = \epsilon_n \psi_n,
\label{NewMehr}
\end{eqnarray}
where ${\bf C}$ is a non-compact discretization of $-\nabla^2$ of the same
order as ${\bf A}_{mehr}$. The solutions of this equation, the $\psi_n$ and
$\epsilon_n$, are the solutions of the original equation. Because ${\bf
A}_{mehr}$ and ${\bf B}_{mehr}$ are finite-differencing operators with
constant coefficients, they are translationally invariant and thus
commute. They are also hermitian. Thus, ${\bf C} = ({\bf B}_{mehr}^{-1} {\bf
A}_{mehr})$ is hermitian, and the wavefunctions of Eq.\
(\ref{Compact-Implicit}) are orthogonal.

Eq.\ (\ref{Compact-Implicit}) is the preferred discretization for computation,
and the equivalent Eq.\ (\ref{NewMehr}) is of formal interest only because the
operators ${\bf B}_{mehr}^{-1}$ and hence ${\bf C}$ are long-ranged and therefore
computationally expensive to apply.

Finally, we consider the invertibility of the ${\bf B}_{mehr}$ operator. We
show that under reasonable conditions ${\bf B}_{mehr}$ has no {\em zero\/}
eigenvalues (in fact, it is a positive definite operator) by arguing that its
null space is empty. It is straightforward to show that the null space of
${\bf B}_{mehr}$ is comprised only of plane waves of maximum kinetic energy;
namely, $\psi_{null}(x, y, z) = e^{-i \pi (x / h_1 + y / h_2 + z / h_3)}$ (or
$y \to -y$, etc.); see Eq.\ (\ref{BofG}) below. Thus, the null space of ${\bf
B}_{mehr}$ is empty whenever these plane waves cannot be represented on the
real-space mesh.

This condition can be realized in two ways: choice of grid size, or explicit
projection. For periodic boundary conditions, when one or more of the linear
dimensions $N_x$, $N_y$, or $N_z$ is odd, the maximum g-vector along that
dimension is $\pi/h_1 (N_x - 1)/N_x < \pi/h_1$. Second, if the grid discretization
cannot be chosen to meet the formal invertibility condition, the pseudo
inverse\cite{GolubVanLoan2} of ${\bf B}_{mehr}$ exists and can be used; that
is, the few vectors in the null space of ${\bf B}_{mehr}$ are projected out
from the wavefunctions. On physical grounds any orbital of such rapid
variation should be excluded from the calculation because it is marginally
representable on the mesh. The pseudo inverse of ${\bf B}_{mehr}$ is
\begin{equation}
{\bf B}_{mehr}^{-1}({\bf x}) =
      \sum_{{\bf g} \ne {\bf g}_{null}}
     e^{-i {\bf x} \cdot {\bf g} } / {\bf B}_{mehr}({\bf g}),
\label{PseudoInverse}
\end{equation}
where the discrete Fourier transform of ${\bf B}_{mehr}$ is
\begin{equation}
{\bf B}({\bf g}) = {1 \over N_{grid}} \sum_{i=1}^3 \cos(h_i {\bf g}_i)^2/3.
\label{BofG}
\end{equation}

\section{Fourier Filtering of Pseudo\-potentials}
The pseudo\-potentials are short-ranged: the Coulomb tail of the local
potential is explicitly canceled and added to the Madelung summation of the
electrostatic energy, and by construction the nonlocal projectors have no
Coulomb tail.  The nonlocal projectors and short-ranged local
pseudo\-potentials are Fourier filtered only once, when the appropriate
potentials and grid spacing are selected.  The filtering procedure attenuates
the high-frequency components while maintaining the localization of the
projectors and potentials.

The unfiltered potentials or projectors are defined on a real-space radial
grid, and are transformed to momentum space in order to filter the
high-frequency components:
\begin{equation}
V_{l,filtered}(G) = F_{filter}(G/G_{cut}) \int V_l(r) j_l(G r) r^2 \, dr,
\end{equation}
where the cut-off function $F_{filter}(G/G_{cut})$ smoothly attenuates the
radial Fourier transform beyond $G > G_{cut}$.  The cut-off wave vector is
determined by the grid spacing: $G_{cut} = \alpha \pi / h_{grid}$.  The
cut-off function is unity for $G < G_{cut}$ and equals $e^{-\beta_1 (G /
G_{cut} - 1)^2}$ for $G > G_{cut}$.  The parameters $\alpha$ and $\beta_1$
depend on the atomic species and are carefully adjusted to achieve the best
results.

After the momentum-space filtering, the back-transformed potentials and
projectors will extend beyond the original core radius.  For computational
efficiency, it is important that the nonlocal pseudo\-potentials be
short-ranged.
Accordingly, a second filtering in real space is applied to reduce the
large-radius oscillations beyond an empirically-determined radius $r_{cut}$.
The second filtering function is unity below the cut-off radius and equals
$e^{-\beta_2 (r/r_{cut} - 1)^2}$ above it.  Example values for a carbon
generalized norm-conserving pseudo\-potential with $s$ and $p$ nonlocalities
are $\alpha = 4/7$ and $\beta_1 = 18$, $r_{cut} = 2.5$ bohr, and $\beta_2 =
0.4$.

Since the filtering procedure modifies the pseudo\-potentials, it is necessary
to determine whether the modified potentials affect the system's physical
properties. Because the degree of filtering is set by the real-space grid
spacing $h_{grid}$, the effect is similar to performing an under-converged
plane-wave calculation. The last effects are well understood and can be
measured quantitatively by progressively increasing the plane-wave cut-off. In
particular, the main results of plane-wave calculations remain
valid, even if they are significantly underconverged.
This is due to the uniform convergence properties of plane waves, which
form a translationally invariant basis set.  Similarly, the convergence
effects may be monitored for a real-space calculation by decreasing the grid
spacing. In our tests we found that the total
energy of the system converges to an asymptotic value in a manner
similar to that observed with plane waves.

\section{Hexagonal Discretization of the Kohn-Sham Equations}

The hexagonal grid structure described in Eq.\ (\ref{HexGrid}) is a simple
hexagonal lattice. Because the $z$ axis is orthogonal to the $xy$ plane, the
$-\nabla^2$ operator may be written in separable form
\begin{equation}
-\nabla^2 = -\nabla^2_{xy} - \nabla^2_z.
\end{equation}
Along the $z$ direction a sixth-order central finite-difference operator was
selected
\begin{equation}
-\nabla^2_z f(i, j, k) =
	 \sum_{n=-3}^3 \alpha_n f(i, j, k+n) + O(h_z^6),
\end{equation}
where $\alpha_{-n} = \alpha_n$ and the $\alpha_n$ are given in Table
\ref{HexOperatorCoefficients}. For the $xy$ plane the the lattice
translational vectors are not orthogonal, and a central finite-difference
expression is not applicable. Instead a composite form was selected
\begin{eqnarray}
-\nabla^2_{xy} f(i, j, k) = &&
	 \sum_{n=-3}^3 \beta_n [ f(i+n, j, k) + f(i, j+n, k)\nonumber\\
        && + f(i+n, j-n, k)] + O(h_{xy}^6),
\end{eqnarray}
where $\beta_{-n} = \beta_n$ and the $\beta_n$ are given in Table
\ref{HexOperatorCoefficients}.

In the multigrid solution process these sixth-order operators are only used on
the finest grid level to compute the kinetic energy and the residual On
coarser grid levels, a second-order operator is used to represent $-\nabla^2$;
viz.,
\begin{equation}
-\nabla^2_z f(i, j, k) =
	 \sum_{n=-1}^1 \alpha'_n f(i, j, k+n) + O(h_z^2),
\end{equation}
and
\begin{eqnarray}
-\nabla^2_{xy} f(i, j, k) = &&
	 \sum_{n=-1}^1 \beta'_n [ f(i+n, j, k) + f(i, j+n, k)\nonumber\\
         &&  + f(i+n, j-n, k)] + O(h_{xy}^2),
\end{eqnarray}
where the discretization weights are listed in Table
\ref{HexOperatorCoefficients}.

The multigrid restriction operator uses a volume weighting scheme with the
weights adjusted for the hexagonal grid, and similarly, the hexagonal
generalization of tri-linear interpolation is used to transfer the coarse-grid
correction to the fine grid. The wavefunctions and Hartree potential are
generated using multigrid iterations in exactly the same manner as was
described in Section III except for the modifications described here.
%

%
\begin{figure}
\caption{Convergence rates for a disordered 64-atom Si cell at a
12 Ry equivalent cut-off. The convergence rate, $\log_{10}(E-E_0)$, is plotted against the number
of self-consistent field (SCF) steps. Random initial wavefunctions were used
with a constant initial density. The initial ionic positions were obtained
from an equilibrated molecular dynamics simulation at 1000 K.  SD represents
convergence rates for the steepest descents algorithm, MG is for multigrid,
SD-SD is steepest descents with subspace diagonalizations, and MG-SD is
multigrid with subspace diagonalizations.}
\label{SiliconConvergence}
\end{figure}
\begin{figure}
\caption{Convergence rates for a 64-atom diamond cell with a substitutional
N impurity at a 63 Ry equivalent cut-off. The convergence rate, $\log_{10}(E-E_0)$, is plotted
against the number of self-consistent field (SCF) steps. Random initial
wavefunctions were used with a constant initial density.  The notation is the
same as in Fig.\ 1.}
\label{DiamondConvergence}
\end{figure}
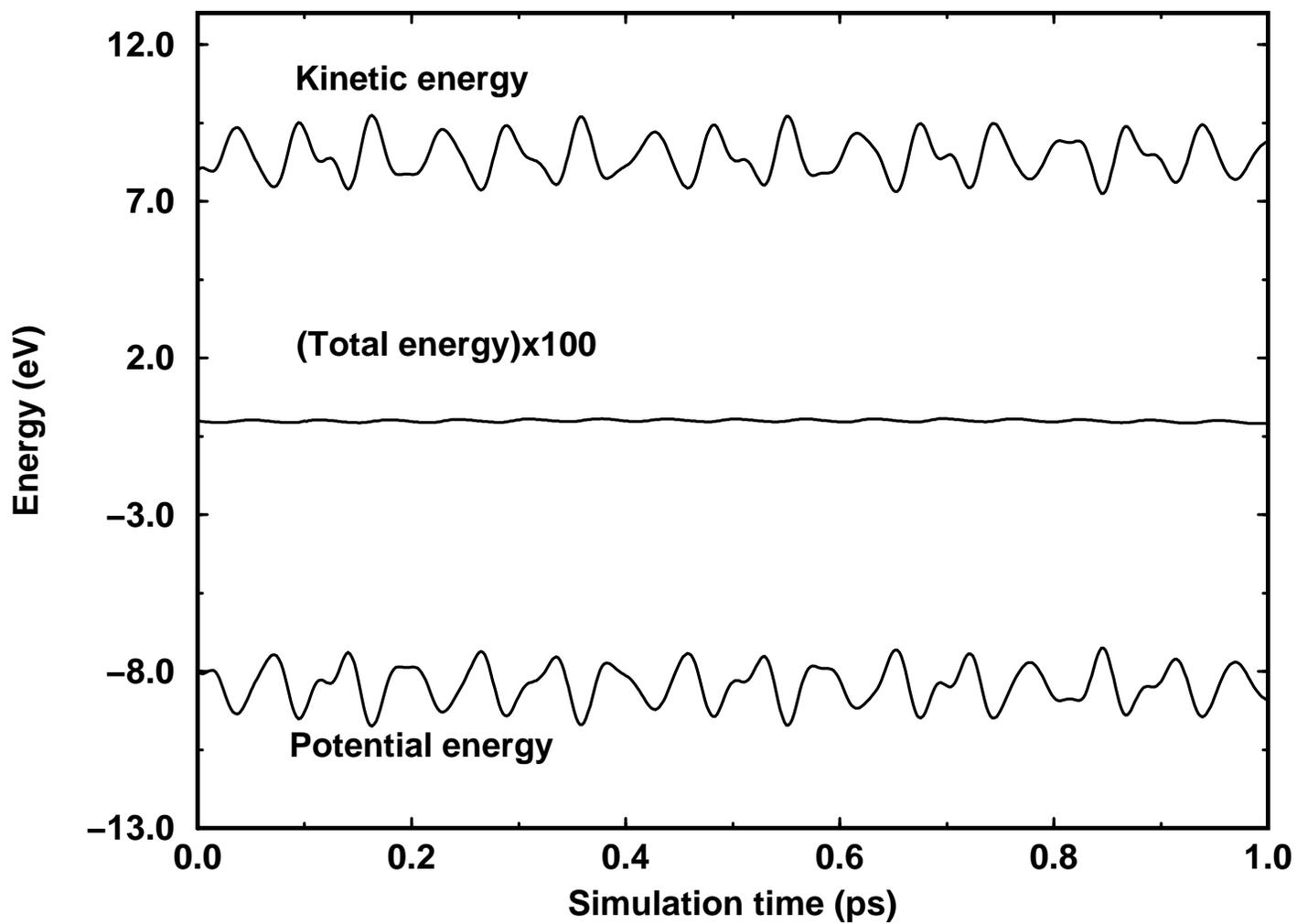
\begin{figure}
\caption{The potential, kinetic, and total energy of a molecular dynamics
simulation of a 64 atom silicon cell at a temperature of 1100 K.  Third-order
Beeman-Verlet integration with an 80 a.u.\ time step was used for the
integration of the ionic equations of motion. The total energy curve is
multiplied by a factor of 100.  The potential and total energies have been
shifted by 251.171 a.u.\ so that they could appear together.}
\label{simd}
\end{figure}
\begin{figure}
\caption{Speedup in execution time is plotted vs.\ number of processors
for a massively parallel implementation of the code on a Cray-T3D.  The test
system is a 64 atom cell of GaN at a fixed, 70 Ry equivalent cut-off.  The
solid line is a guide to the eye.}
\label{speedup}
\end{figure}
%
\begin{table}
\caption{Discretization weights for the fourth-order orthorhombic
{Mehrstellen\/} operators for the central, nearest-neighbor, and next
nearest-neighbor grid points.  See Eqs. (A12) and (A2) for the definitions of
$a$, $b_n$, and $c_{n,m}$.  The cubic-grid operator corresponds to $h_i =
h_{grid}$.}
\begin{tabular}{lccc}
                 & $a$ & $b_n$ & $c_{n,m}$ \\
\tableline
${\bf A}_{mehr}$ & $\sum_i {4 \over 3 h_i^2}$
                 & $-{5 \over 6 h_n^2} + \sum_i {1\over 6h_i^2}$
                 & $-{1 \over 12h_n^2} - {1 \over 12h_m^2} $ \\
${\bf B}_{mehr}$ & ${1 \over  2}$
                 & ${1 \over 12}$
                 & 0
\end{tabular}
\label{Mehrstellen-Ortho}
\end{table}
\begin{table}
\caption{Discretization weights, listed by distance along basis set axes,
for the sixth and second-order kinetic energy operators for the hexagonal
grid.  See also Eqs. C2-C5.}
\begin{tabular}{clcccc}
  operator & order & 0 & 1 & 2 & 3 \\
\tableline
$-\nabla^2_z$ $h_z^2$ & 2$^{\hbox{nd}}$ &     2 &   -1 &     0 &    0 \\
                      & 6$^{\hbox{th}}$ & 49/18 & -3/2 & 3/20 & -1/90 \\
\tableline
$-\nabla^2_{xy}$ $h_{xy}^2$ & 2$^{\hbox{nd}}$ & 4/3 & -2/3 & 0 & 0 \\
                            & 6$^{\hbox{th}}$ & 49/27&-1&1/10&-1/135 \\
\end{tabular}
\label{HexOperatorCoefficients}
\end{table}
\end{document}